\documentclass[pra,aps,twocolumn,epsfig,superscriptaddress,showpacs]{revtex4}
\usepackage{bm}
\usepackage{amsfonts}
\usepackage[dvips]{graphicx}
\usepackage{mathrsfs}
\usepackage[intlimits]{amsmath}
\usepackage[colorlinks=false,dvipdfm]{hyperref}

\begin{document}
\author{Jing-Ling Chen}
 \email{chenjl@nankai.edu.cn}
\affiliation{Theoretical Physics Division, Chern Institute of
Mathematics, Nankai University, Tianjin 300071, People's Republic of
China}
\author{Dong-Ling Deng}
 \affiliation{Theoretical Physics Division, Chern Institute of
Mathematics, Nankai University, Tianjin 300071, People's Republic of
China}

\date{\today}

\title{Bell inequality for qubits based on the Cauchy-Schwarz
inequality}

\begin{abstract}
We develop a systematic approach to establish Bell inequalities for
qubits based on the Cauchy-Schwarz inequality. We also use the
concept of \emph{distinct ``roots"} of Bell function to classify
some well-known Bell inequalities for qubits. As applications of the
approach, we present three new and tight Bell inequalities for four
and three qubits, respectively.
\end{abstract}

 \maketitle

Bell inequality has been regarded as ``the most profound discovery
in science" \cite{Stapp}. It is at the heart of the study of
nonlocality and is the most famous legacy of the late physicist John
S. Bell \cite{J.S.Bell}. The inequality shows that the predictions
of quantum mechanics are not intuitive, and touches upon fundamental
philosophical issues that relate to modern physics. Bell-test
experiments serve to investigate the validity of the entanglement
effect in quantum mechanics by using some kinds of Bell
inequalities, however they to date overwhelmingly show that Bell
inequalities are violated. These experimental results provide
empirical evidence against local realism \cite{A.Einstein} and in
favor of quantum mechanics.


Bell's applaudable progress has stirred a great furor. Many people
have been attracted in this problem and extensive work on Bell
inequalities has been done, including both theoretical analysis and
experimental test. The famous Clauser-Horne-Shimony-Holt
(CHSH)~\cite{J.Clauser} inequality is a kind of improved Bell
inequality that is more convenient for experiments. Now, it is
well-known that all pure entangled states of two two-dimensional
systems (i.e., qubits) violate the CHSH inequality and the maximum
quantum violation is the so-called Tsirelson's bound $2\sqrt{2}$.
Mermin, Ardehali, Belinskii, and Klyshko have separately generalized
the CHSH inequality to the $N$-qubit case, which now known as the
MABK inequality, and proved that quantum violation of this
inequality increases with the number of particles~\cite{MABK}. In
$2001$, Scarani and Gisin~\cite{2001-Scarani} noticed that the
generalized Greenberger-Horne-Zeilinger (GHZ)
states~\cite{GHZ-states}:
$|\psi\rangle_{GHZ}=\cos\xi|0\cdots0\rangle+\sin\xi|1\cdots1\rangle$
do not violate the MABK inequality for
$\sin2\xi\leq1/\sqrt{2^{N-1}}$ (The GHZ state is for $\xi=\pi/4$).
In Refs. \cite{zukowski1,Werner1} a general correlation-function
$N$-qubit Bell inequality has been derived, hereafter we call it as
the Werner-Wolf-{\.Z}ukowski-Brukner (WWZB) inequality. The WWZB
inequality includes the MABK inequality as a special case. Ref.
\cite{zukowski2} shows that (a) For $N={\rm even}$, although the
generalized GHZ state does not violate the MABK inequality, it does
violate the WWZB inequality, and (b) For $\sin2\xi \le
1/\sqrt{2^{N-1}}$ and $N={\rm odd}$, the WWZB inequality cannot yet
be violated for the whole region $\xi \in [0, \pi/2]$. For the
three-qubit case, such a difficulty has been overcome in Ref.
\cite{Chen}, where a probabilistic Bell inequality was proposed and
consequently Gisin's theorem for three qubits naturally returned.
Recent development also indicates that Bell inequality is not unique
when one studies Gisin's theorem for three qubits, in Ref.
\cite{JL-Chen-2008} three of such inequalities have been listed and
compared. All the inequalities mentioned above belong to the
two-setting Bell inequalities for $N$ qubits, i.e., they are based
on the standard Bell experiment, in which each local observer is
given a choice between two dichotomic observables. Some significant
generalizations have been made for multi-setting Bell inequalities
for $N$ qubits \cite{zukowski3} as well as two-setting Bell
inequalities for high-dimensional systems
\cite{2002Collins,Chen2008-1}. Notably, in Ref. \cite{2004W.
Laskowski} a multi-setting Bell inequality that violate the
generalized GHZ state for the whole region $\xi \in [0, \pi/2]$ was
found.

In this paper, we shall focus on Bell inequality for qubits. It is
natural to ask two questions: (i) As an inequality, does Bell
inequality have any deep connections with some ancient mathematic
inequalities, such as the Cauchy-Schwarz inequality or more
generally, the H{\" o}lder inequality? (ii) So far, many kinds of
Bell inequalities for qubits have emerged, even for three qubits.
Can these inequalities be classified in an efficient way? The
purpose of this paper is to (i) develop a systematic approach to
establish Bell inequalities for qubits based on the Cauchy-Schwarz
inequality, and (ii) classify some well-known Bell inequalities
based on their distinct ``roots". Let us start from the weighed H{\"
o}lder inequality since it is a more general inequality which
contains the Cauchy-Schwarz inequality as a special case.



\emph{The Cauchy-Schwarz inequality and  Bell inequalities}. Let
$f(\lambda)$ and $g(\lambda)$ be  any two real functions for which
$|f(\lambda)|^p$ and $|g(\lambda)|^q$ are integrable in $\Gamma$
with $p> 1$ and $\frac{1}{p}+\frac{1}{q}=1$, then the weighed H{\"
o}lder inequality reads \cite{I.S.Gradshteyn-holder}:
\begin{eqnarray}\label{Holder}
\int_{\Gamma} f g \rho(\lambda)
d\lambda\leq\left[\;\int_{\Gamma}|f|^p\rho(\lambda)
d\lambda\right]^\frac{1}{p} \left[\;\int_{\Gamma}|g|^q\rho(\lambda)
d\lambda\right]^\frac{1}{q},
\end{eqnarray}
where $\Gamma$ is the total $\lambda$ space and $\rho(\lambda)$ is a
statistical distribution of $\lambda$, which satisfies
$\rho(\lambda)\geq0$ and $\int_\Gamma d\lambda\rho(\lambda)=1$. When
$p=q=2$, the above inequality reduces to the Cauchy-Schwarz
inequality: $\left[\int_{\Gamma}
fg\rho(\lambda)d\lambda\right]^2\leq\left[\int_{\Gamma}|f|^2\rho(\lambda)d\lambda\right]
\left[\int_{\Gamma}|g|^2\rho(\lambda)d\lambda\right]$. It may be
very interesting to consider the cases when $p$, $q$ take various
values. However, in this paper, we restrict our study to the case
with $p=q=2$, which is easier for the calculations.

Consider $N$ spatially separated parties and allow each of them to
choose independently among $M$ observables, determined by some local
parameters denoted by $\lambda$ . Let
$X_j(\mathbf{n}_{k_j},\lambda)$, or $X_{j,k_j}$ for simplicity,
denote observables on the $j$-th qubit, each of which has two
possible outcomes $-1$ and $1$. From the viewpoint of local realism,
the values of $X_j$'s are predetermined by the local hidden variable
(LHV) $\lambda$ before measurement, and independent of any
measurements, orientations or actions performed on the other parties
at spacelike separation. The correlation function, in the case of a
local realistic theory, is then defined as
$Q(\mathbf{n}_{k_1},\mathbf{n}_{k_2},\cdots,\mathbf{n}_{k_j})=\int_\Gamma
\Pi_{j=1}^NX_j(\mathbf{n}_{k_j},\lambda) \rho(\lambda) d\lambda$,
where $j=1,2,\cdots,N$ and $k_j=1,2,\cdots,M$. For convenience, we
denote the correlation function
$Q(\mathbf{n}_{k_1},\mathbf{n}_{k_2},\cdots,\mathbf{n}_{k_j})$ as
$Q_{k_1k_2\cdots k_j}$. In the following, we present a systematic
approach to establish Bell inequalities for qubits based on the
Cauchy-Schwarz inequality. It contains three main steps:

\emph{Step 1:}  {\it Connecting functions $f(\lambda)$ and
$g(\lambda)$ with the observables}. $f(\lambda)$ and $g(\lambda)$ in
inequality~(\ref{Holder}) can be functions of the observables of the
parties. To express the functions more concisely, let $X_{j,0}=1$,
then one has
\begin{eqnarray}\label{general-f-g}
f(\lambda)=\sum_{\chi}C_{\chi}\prod_{j=1}^{N}X_{j,k_j},\;\;\;
g(\lambda)=\sum_{\chi}D_{\chi}\prod_{j=1}^{N}X_{j,k_j}.
\end{eqnarray}
Here we associate to each observables, $X_{1,1}$, $X_{1,1}X_{2,1}$,
or generally $\prod_{j=1}^{N}X_{j,k_j}$, a single symbol $\chi$,
which stands for $N$ pairs of indices (one pair for each observer).
Obviously, there are $N_\chi=(1+M)^N$ distinct values of $\chi$. The
constant numbers $C_{\chi}$ and $D_{\chi}$ are coefficients of
$\prod_{j=1}^{N}X_{j,k_j}$.

\emph{Step 2}: {\it Establishing Bell inequalities by determining
coefficients $C_{\chi}$ and $D_{\chi}$}. Note that for the qubit
case, one always has $X^2_{j,k_j}=1$, which is useful for
simplifying $f(\lambda) g(\lambda)$, $f^2(\lambda)$ and
$g^2(\lambda)$ in the inequality~(\ref{Holder}). One will find that
some terms in $f(\lambda) g(\lambda)$, $f^2(\lambda)$ and
$g^2(\lambda)$, such as $X_{1,1}X_{1,2}$, $X_{1,1}X_{1,2}X_{2,1}$,
or $X_{1,1}X_{1,2}X_{2,1}X_{2,2}$, etc., are impossible to calculate
in quantum mechanics. Consequently, we shall set all the
coefficients of such terms be zero, and then get a series of
equations for $C_{\chi}$ and $D_{\chi}$. By solving these equations,
we obtain the solutions of $C_{\chi}$ and $D_{\chi}$. Substituting
these solutions into Eqs. (\ref{Holder}) (\ref{general-f-g}), one
gets a set of Bell inequalities: $\langle
f(\lambda)g(\lambda)\rangle_{LHV}^2\leq\langle
f(\lambda)\rangle_{LHV}^2 \langle g(\lambda)\rangle_{LHV}^2$. Here
$\langle f(\lambda)g(\lambda)\rangle_{LHV}=\int_\Gamma
f(\lambda)g(\lambda) \rho(\lambda) d\lambda$, and $\langle
f(\lambda)\rangle_{LHV}^2$, $\langle g(\lambda) \rangle_{LHV}^2$ are
defined similarly.

\emph{Step 3}: {\it Ruling out the trivial Bell inequalities}. Some
inequalities obtained in \emph{Step 2} are trivial, i.e., they
cannot be violated in quantum mechanics. Thus we should rule them
out by calculating the quantum violation of each inequality. In this
step, we finally achieve some nontrivial Bell inequalities, such as
the tight inequalities.

Here we present an example to illuminate this method.

\emph{Example 1}: {\it Derivation of the CHSH inequality}. Let us
look at the simplest case, i.e., $N=2,M=2$. Let
\begin{eqnarray}\label{2qubits-f-g}
f(\lambda)&=&C_0+C_1[X_{1,1}+X_{2,1}]+C_2[X_{1,2}+
X_{2,2}]\nonumber\\
&&+C_3X_{1,1}X_{2,1}+C_4[X_{1,1}X_{2,2}+X_{1,2}X_{2,1}]\nonumber\\ &&+C_5X_{1,2}X_{2,2},\nonumber\\
g(\lambda)&=&D_0+D_1[X_{1,1}+X_{2,1}]+D_2[X_{1,2}+
X_{2,2}]\nonumber\\&&+D_3X_{1,1}X_{2,1}
+D_4[X_{1,1}X_{2,2}+X_{1,2}X_{2,1}]\nonumber\\&& +D_5X_{1,2}X_{2,2},
\end{eqnarray}
then we get a series of equations of $C_j$'s and $D_j$'s (here we
omit them for sententiousness). After solving these equations, we
finally choose the solutions: $C_0$ and $D_0$ are arbitrary real
numbers, $C_1=C_2=0, C_3=C_4=-C_5$, $D_1=D_2=0, D_3=D_4=-D_5$, then
from $\langle f(\lambda)g(\lambda)\rangle^2\leq\langle
f(\lambda)\rangle^2\langle\; g(\lambda)\rangle^2$ we have a Bell
inequality as:
$(Q_{11} + Q_{12} + Q_{21} - Q_{22})^2\leq 4 $, or $(Q_{11} + Q_{12}
+ Q_{21} - Q_{22})/2\leq 1$.
It is nothing but the famous CHSH inequality!


Similarly, the MABK and the WWZB inequalities can also be derived
with the same approach, although the computation becomes a bit more
complicated when $N$ increases. However, the above approach can be
improved further with the aid of the distinct ``roots" of the Bell
function. What do we mean the ``roots" of the Bell function? For
instance, let the left-hand side of the CHSH inequality ${\cal
B}(\lambda)
=(X_{1,1}X_{2,1}+X_{1,1}X_{2,2}+X_{1,2}X_{2,1}-X_{1,2}X_{2,2})/2$ be
the Bell function,
for each set of values [such as $\{X_{1,1}=1, X_{1,2}=1, X_{2,1}=1,
X_{2,2}=1 \}$], the Bell function corresponds to a number [such as
${\cal B}(\lambda)=1$]. This number is called a ``root" of the Bell
function ${\cal B}(\lambda)$. Obviously, there are totally $2^4=16$
sets of values $\{X_{1,1}, X_{1,2}, X_{2,1}, X_{2,2}\}$, so ${\cal
B}(\lambda)$ has totally 16 roots. However, 8 roots equal to $-1$,
the other 8 roots equal to 1, therefore ${\cal B}(\lambda)$ has only
two distinct ``roots": $\Lambda_1=-1$ and $\Lambda_2=1$. Then for
any set of values $\{X_{1,1}, X_{1,2}, X_{2,1}, X_{2,2}\}$, one
always has the algebraic equation: $[{\cal
B}(\lambda)-\Lambda_1][{\cal B}(\lambda)-\Lambda_2]=0$, or ${\cal
B}^2(\lambda)=1$. We have the following Theorem.

\emph{Theorem 1.}  Let $S_2=\{\mathcal {B}(\lambda) \;|\; \mathcal
{B}^2(\lambda)=1\}$, for $\forall \; \mathcal{B}(\lambda)\in S_2$,
we have Bell inequality ${\cal I}=|\langle
\mathcal{B}(\lambda)\rangle_{LHV}|\leq 1$.

\emph{Proof.} Let $f(\lambda)=1+\mathcal {B}(\lambda)$,
$g(\lambda)=1-\mathcal {B}(\lambda)$, then from the Cauchy-Schwarz
inequality we have $[\int[1+\mathcal {B}(\lambda)][1-\mathcal
{B}(\lambda))\rho(\lambda)d\lambda]\leq\int[1+\mathcal
{B}(\lambda)]^2 \rho(\lambda)d\lambda\int[1-\mathcal
{B}(\lambda)]^2\rho(\lambda)d\lambda$, which yields $\langle
\mathcal {B}(\lambda)\rangle^2_{LHV} \le \langle \mathcal
{B}^2(\lambda)\rangle_{LHV}$. Because $\mathcal {B}(\lambda)\in
S_2$, which implies ${\mathcal {B}}^2(\lambda)=1$, thus we arrive at
the Bell inequality $|\langle \mathcal
{B}(\lambda)\rangle_{LHV}|\leq 1$. This ends the proof.

Based on \emph{Theorem 1}, there is a simpler way to derive Bell
inequality as follows: Let
\begin{eqnarray}\label{B}
\mathcal {B}(\lambda)= \frac{1}{2}\biggr[{X}(\lambda)+{Y}(\lambda)+
{Z}(\lambda)- {X}(\lambda) {Y}(\lambda) {Z}(\lambda)\biggr],
\end{eqnarray}
one easily proves that $\mathcal {B}^2(\lambda)=1$, provided
${X}^2(\lambda)={Y}^2(\lambda)={Z}^2(\lambda)=1$, then from
\emph{Theorem 1} one has a Bell inequality ${\cal I}=|\langle
\mathcal{B}(\lambda)\rangle_{LHV}|\leq 1$. For instance, let
${X}(\lambda)={A_1}(\lambda){B_1}(\lambda)$,
${Y}(\lambda)={A_1}(\lambda){B_2}(\lambda)$,
${Z}(\lambda)={A_2}(\lambda){B_1}(\lambda)$, which yields
${X}(\lambda) {Y}(\lambda)
{Z}(\lambda)={A_2}(\lambda){B_2}(\lambda)$ [using
${A_j}^2(\lambda)=1$, ${B_j}^2(\lambda)=1$], then one has the CHSH
inequality: ${\cal I}_{CHSH}=|\langle
\mathcal{B}(\lambda)\rangle_{LHV}|=  |\langle A_1 B_1+A_1
B_2+A_2B_1-A_2B_2 \rangle/2| \leq 1$; Let
${X}(\lambda)={A_1}(\lambda){B_1}(\lambda){C_2}(\lambda)$,
${Y}(\lambda)={A_1}(\lambda){B_2}(\lambda){C_1}(\lambda)$,
${Z}(\lambda)={A_2}(\lambda){B_1}(\lambda){C_1}(\lambda)$, one has
the MABK inequality for three qubits as ${\cal I}_{MABK}=|\langle
\mathcal{B}(\lambda)\rangle_{LHV}|=  |\langle A_1 B_1 C_2+A_1 B_2
C_1+ A_2 B_1 C_1-A_2 B_2 C_2 \rangle/2| \leq 1$.

The Bell functions of the MABK and the WWZB inequalities for $N$
qubits have two distinct ``roots" 1 and $-1$, so they belong to
$S_2$. The Bell function $\mathcal {B}(\lambda)$ in Eq. (\ref{B})
naturally connects with the MABK inequality or the WWZB  inequality
in the following way: Let ${X}(\lambda)=\mathcal {B}_{N-1}(\lambda)
X_{N, 1}$, ${Y}(\lambda)=\mathcal {B}_{N-1}(\lambda) X_{N, 2}$,
${Z}(\lambda)=\mathcal {B}'_{N-1}(\lambda) X_{N, 1}$, where
$\mathcal {B}_{N-1}(\lambda)$ is the Bell function of MABK
inequality or the WWZB inequality for $(N-1)$ qubits, and $\mathcal
{B}'_{N-1}(\lambda)$ is obtained through the interchanges
$X_{j,1}\leftrightarrow X_{j, 2}$. After substituting them into Eq.
(\ref{B}) and using ${X}(\lambda) {Y}(\lambda) {Z}(\lambda)=\mathcal
{B}'_{N-1}(\lambda) X_{N, 2}$, one has $\mathcal
{B}(\lambda)=\frac{1}{2}\{\mathcal {B}_{N-1}(\lambda) [X_{N,
1}+X_{N, 2}]+\mathcal {B}'_{N-1}(\lambda) [X_{N, 1}-X_{N, 2}]\}$,
which is just the Bell function of MABK and WWZB inequalities. By
the way, if $\mathcal {B}'_{N-1}(\lambda)$ is replaced by identity
${\bf 1}_{N-1}=\prod_{j=1}^{N-1}X_{j,0}$, then one recovers the
family of two-setting Bell inequality for many qubits as shown in
Ref. \cite{K.Chen} (see inequality (4) in \cite{K.Chen}). Also, both
the three-setting Bell inequality (28) and the four-setting Bell
inequality (35) presented in Ref. \cite{Marcin} belong to $S_2$.

For Bell functions with three or more distinct ``roots", we have the
following Theorem.

\emph{Theorem 2.}  Let $S_3=\{\mathcal {B}(\lambda) \;|\; \mathcal
{B}^3(\lambda)=\mathcal {B}(\lambda), \; \mathcal {B}(\lambda)\notin
S_2\}$, i.e., $\mathcal {B}(\lambda)$ must have three distinct
``roots" $\Lambda_1=-1$, $\Lambda_2=0$, and $\Lambda_3=1$, then for
$\forall \; \mathcal {B} (\lambda)\in S_3$, one has the Bell
inequality: $|\langle \mathcal {B} (\lambda)\rangle_{LHV}|\leq 1$.
In general, if  $ S_n=\{\mathcal {B}(\lambda) \;|\;
\prod_{j=0}^{j=n-1} (\mathcal {B}-\Lambda_j)=0,\;\mathcal
{B}(\lambda)\notin \bigcup_{k=2}^{k=n-1}S_k, \; n\in {\rm integers},
\; n \ge 3\}$, which means that $n$ ``roots" of $\mathcal
{B}(\lambda)$ uniformly distribute between $-1$ and $1$ with
$\Lambda_j=-1+2j/(n-1)$, for $\forall \; \mathcal {B} (\lambda)\in
S_n$, one has the Bell inequality: $|\langle \mathcal {B}
(\lambda)\rangle_{LHV}|\leq 1$.

\emph{Proof.} First, we prove that for $n=3$ the theorem is valid.
Since $\langle \mathcal {B}(\lambda)\rangle^2 \le \langle \mathcal
{B}^2(\lambda)\rangle$, what we need to do is to prove $\langle
\mathcal {B}^2(\lambda)\rangle \le 1$. Let $f(\lambda)=1+\mathcal
{B}^2(\lambda)$, $g(\lambda)=1-\mathcal {B}^2(\lambda)$, from the
Cauchy-Schwarz inequality one obtains $\langle \mathcal
{B}^2(\lambda)\rangle^2 \le \langle \mathcal {B}^4(\lambda)\rangle$.
By using $\mathcal {B}(\lambda)^3=\mathcal {B}(\lambda)$, we have
$\langle \mathcal {B}^2(\lambda)\rangle^2 \le \langle \mathcal
{B}^4(\lambda)\rangle=\langle \mathcal {B}^2(\lambda)\rangle$, i.e.,
$\langle \mathcal {B}^2(\lambda)\rangle [\langle \mathcal
{B}^2(\lambda)\rangle-1]\le 1$. Because $\langle \mathcal
{B}^2(\lambda)\rangle \ge 0$, then we have $\langle \mathcal
{B}^2(\lambda)\rangle_{LHV}\leq1$. Second, we use the induction
method to prove this theorem. Suppose for all $n\leq k $, the
theorem is valid, then for $\mathcal{B}_{k+1}\in S_{k+1}$, one has
$\mathcal{B}_{k+1}=(\mathcal{B}_{k+1}^2-1)\prod_{j=1}^{j=k-1}(\mathcal{B}_{k+1}-\Lambda_j)=0$.
If $(\mathcal{B}_{k+1}^2-1)=0$, \emph{Theorem 1} yields $|\langle
\mathcal{B}_{k+1}\rangle|\leq 1$. If
$\prod_{j=1}^{j=k-1}(\mathcal{B}_{k+1}-\Lambda_j)=0$, from the
principle of induction method, one obtains
$|\langle\mathcal{B}_{k+1}\rangle|=|\langle\frac{k-2}{k}\mathcal{B}_{k-1}\rangle|\leq
\frac{k-2}{k}$. Thus, one has $|\langle \mathcal {B}
(\lambda)\rangle_{LHV}|\leq 1$, which completes the proof.

Let us see what Bell inequalities appeared in literature belong to
$S_3$. The first example is the three-setting Bell inequality for
$N$ qubits proposed in Ref. \cite{M.Zukowski} [see inequality (9) in
\cite{M.Zukowski} or inequality (2) in \cite{JL-Chen-2002}].
Although this inequality is not tight, it has a particular property
that for the GHZ state it is more resistant to noise than the MABK
inequality when $N \ge 4$. The second example is the two two-setting
Bell inequalities listed in Ref. \cite{JL-Chen-2008} [see
inequalities (3) and (4) in \cite{JL-Chen-2008}]; These two
inequalities are not tight but they are violated by any pure
entangled state of three qubits. Moreover, there indeed exist some
Bell inequalities belonging to $S_4$. The first example is a
three-setting Bell inequality for two qubits proposed in Ref.
\cite{D. Collins-N.Gisin} [see inequality (19) in \cite{D.
Collins-N.Gisin}]. It is a relevant two-qubit Bell inequality
inequivalent to the CHSH inequality. The most interesting feature of
this tight inequality is that there exist states that violate it,
but do not violate the CHSH inequality. The second example is the
two-setting Bell inequality presented in Ref. \cite{JL-Chen-2008}
[see inequality (8) in \cite{JL-Chen-2008}]. It is a tight Bell
inequality and is violated by any pure entangled state of three
qubits. To our knowledge, Bell inequalities belonging to $S_n$
$(n\ge 5)$ have seldom appeared in the literature.

As applications of the approach, we present three new and tight Bell
inequalities that belong to $S_3$ as follows.

\emph{Example 2}: {\it A new two-setting Bell inequality for four
qubits $ \langle I_{4,2} \rangle\le 1$.} Let $\mathcal
{B}(\lambda)=\frac{7}{16}+\frac{9 }{16}I_{4,2}(\lambda)$, where
$I_{4,2}=(1/9)(-5Q_{1111}-2Q_{2222}-[Q_{1120}+Q_{1210}+Q_{2110}+Q_{1102}]
-[Q_{2200}+Q_{2020}+Q_{2002}+Q_{0220}+Q_{0202}+Q_{0022}]
+[Q_{2000}+Q_{0200}+Q_{0020}+Q_{0002}]+2[Q_{1000}+Q_{0100}+Q_{0010}+Q_{0001}]
+[Q_{1110}+Q_{1101}+Q_{1011}+Q_{0111}]
+[Q_{1201}+Q_{2101}+Q_{1012}+Q_{1102}+Q_{2011}+Q_{0112}+Q_{0121}
+Q_{0211}]+2[Q_{1112}+Q_{1121}+Q_{1211}+Q_{2111}]
-[Q_{1220}+Q_{2120}+Q_{2210}+Q_{1202}+Q_{2102}+Q_{2201}+Q_{1022}
+Q_{2021}+Q_{2012}+Q_{0122}+Q_{0221}+Q_{0212}]
+[Q_{1122}+Q_{1212}+Q_{1221}+Q_{2211}+Q_{2121}+Q_{2112}])$,
$Q_{mnkl}=\int_{\Gamma} X_{1,m}X_{2,n}X_{3,k}X_{4,l}
\rho(\lambda)d\lambda$, $(m,n,k,l=0,1,2)$, are the correlation
functions, and here we have set $X_{j,0}=1$ ($j=1,2,3,4$). It is
easy to check $\mathcal {B}(\lambda) \in S_3$, therefore from
\emph{Theorem 2} we have the Bell inequality $|\langle \mathcal {B}
(\lambda)\rangle_{LHV}|\leq 1$, or $ \langle I_{4,2} \rangle\le 1$.
This inequality $ \langle I_{4,2} \rangle\le 1$ is tight and is
symmetric under the permutations of $X_{j,k}$'s.

For four qubits, the Werner states read
$\rho_W=V|\psi\rangle\langle\psi|+(1-V)\rho_{noise}$, where
$|\psi\rangle=(1/\sqrt{2})(|0000\rangle+|1111\rangle)$ is the
four-qubit GHZ state, $V$ is the so-called visibility,
$\rho_{noise}=\mathbb{I}_{16}/16$, and $\mathbb{I}_{16}$ is the
$16\times 16$ identity matrix. In quantum mechanics, the observables
read $X_{j,k}=\vec{\sigma}\cdot\vec{n}_{k_j}$, $(j=1, 2, 3, 4; k=1,
2)$, $\vec{\sigma}$ is the vector of Pauli matrices,
$\vec{n}_{k_j}=(\sin\theta_{k_j} \cos\phi_{k_j}, \sin\theta_{k_j}
\sin\phi_{k_j},\cos\theta_{k_j})$, $X_{j,0}=\mathbb{I}_2$ is the $2
\times 2$ identity matrix.
The correlation functions for the Werner states read $Q_{mnkl}={\rm
Tr}(\rho_{W} \; X_{1,m}\otimes X_{2,n}\otimes X_{3,k}\otimes
X_{4,l})$.
For the GHZ state, the threshold visibility is
$V_{GHZ}^{4-qubit}\approx9/15.56 \approx0.5784$, which is larger
than $1/2\sqrt{2}$, the result given by the Mermin inequality.
Namely, the inequality $ \langle I_{4,2} \rangle\le 1$ is less
resistant to noise. However, numerical result shows that the
generalized GHZ states violate $\langle I_{4,2} \rangle\le 1$ for
the whole region, which cannot be done for the MABK inequality. By
setting $X_{4,1}=X_{4,2}=1$, $\langle I_{4,2} \rangle\le 1$ reduces
to an equivalent form of the three-qubit Bell inequality (4) in Ref.
\cite{JL-Chen-2008}. Remarkably, numerical calculation indicates
that Bell inequality $ \langle I_{4,2} \rangle\le 1$ is violated by
all pure entangled states of four qubits, thus it is a good
candidate for proving Gisin's theorem for four qubits.

\emph{Example 3}: {\it Another new two-setting Bell inequality for
four qubits $ \langle I'_{4,2} \rangle\le 1$.} Let $\mathcal
{B}(\lambda)=\frac{6}{16}+\frac{10}{16}I'_{4,2}(\lambda)$, where
$I'_{4,2}=(1/10)(-[Q_{1200}+Q_{2100}+Q_{1020}+Q_{2010}+Q_{1002}+Q_{2001}
+Q_{0120}+Q_{0210}+Q_{0102}+Q_{0201}+Q_{0012}+Q_{0021}]-Q_{2222}
+[Q_{1112}+Q_{1121}+Q_{1211}+Q_{2111}]+3[Q_{1000}+Q_{0100}+Q_{0010}
+Q_{0001}]+[Q_{2000}+Q_{0200}+Q_{0020}+Q_{0002}]
-[Q_{1220}+Q_{2120}+Q_{2210}+Q_{1202}+Q_{2102}+Q_{2201}+Q_{1022}
+Q_{2021}+Q_{2012}+Q_{0122}+Q_{0221}+Q_{0212}]
+[Q_{1122}+Q_{1212}+Q_{1221}+Q_{2211}+Q_{2121}+Q_{2112}]
-[Q_{2220}+Q_{2202}+Q_{2022}+Q_{0222}]-3Q_{1111})$. One may verify
that $\mathcal {B}(\lambda) \in S_3$, thus we have the Bell
inequality $ \langle I'_{4,2} \rangle\le 1$. This inequality is
tight and is violated by the generalized GHZ states for the whole
region. It is also a good candidate for proving the Gisin's theorem
for four qubits.


\emph{Example 4}: {\it A new three-setting Bell inequality for three
qubits $ \langle I_{3,3} \rangle\le 1$}. The Bell inequality reads
\begin{eqnarray}\label{3settings-3qubits}
\langle I_{3,3} \rangle &=&[(Q_{223}+Q_{232}+Q_{322})-2(Q_{211}+Q_{121}+Q_{112})\nonumber\\
&&+(Q_{221}+Q_{122}+Q_{212})-(Q_{331}+Q_{313}+Q_{133})\nonumber\\
&&+(Q_{321}+Q_{312}+Q_{213}+Q_{231}+Q_{123}+Q_{132})\nonumber\\
&&+2Q_{111}+4Q_{222}-Q_{333}]/8 \leq 1,
\end{eqnarray}
and $\mathcal {B}(\lambda) =I_{3,3}(\lambda)\in S_3$. This
inequality is also tight, which reduces to the CHSH inequality when
$X_{3,1}=1$, $X_{3,2}=X_{3,3}=-1$. It is resistant to noise as
strong as the MABK inequality for the GHZ state (i.e., $\langle
I_{3,3}^{max}\rangle=2$), but is violated by the generalized GHZ
states for a wider region (for instance, $\langle
I_{3,3}^{max}\rangle \approx1.0059$ when $\xi=\pi/12$).


In conclusion, we have developed a systematic approach to establish
Bell inequalities for qubits based on the Cauchy-Schwarz inequality.
We have also used the concept of \emph{distinct ``roots"} of Bell
function to classify some well-known Bell inequalities for qubits.
As applications of the approach, we have presented three new and
tight Bell inequalities. In addition, there is an alternative way to
derive the CHSH inequality: From \emph{Theorem 1} and Eq. (\ref{B}),
one may have $ {\cal I}_{1}=\langle 1+ A_1 B_1 -A_2 B_2 + A_1 A_2
B_1 B_2 \rangle /2 \le 1$, ${\cal I}_{2}= \langle 1+ A_1 B_2 +A_2
B_1 - A_1 A_2 B_1 B_2 \rangle /2 \le 1.$
By adding up these two inequalities, one arrives at the famous CHSH
inequality: ${\cal I}_{CHSH}=  {\cal I}_{1}+ {\cal I}_{2}=\langle
A_1 B_1+A_1 B_2+A_2B_1-A_2B_2 \rangle/2 \leq 1$. Usually, combining
two inequalities directly will lead to an inequality with looser
constraint than before. In such a sense, inequalities ${\cal I}_{1}$
and ${\cal I}_{2}$ are stronger than the CHSH inequality, but the
difficult point is that it is impossible to compute the term
$\langle A_1 A_2 B_1 B_2 \rangle$ for quantum mechanics. Some time
ago, Gisin posed a question to find Bell inequalities which are more
efficient than the CHSH one for Werner states \cite{N.Gisin1} (see
also \cite{N.Gisin2}). Recently, V{\' e}rtesi has given a positive
answer to Gisin's question by providing a new family of
multi-setting (at least $M=465$ settings for each party) Bell
inequalities \cite{Vertesi}, which proves the two-qubit Werner
states to be nonlocal for a wider parameter range $0.7056 < V \le
1$.
How to use \emph{Theorems 1} and \emph{2} to construct even more
efficient Bell inequalities for Werner states is a significant
topic, which we shall investigate subsequently.

This work is supported in part by NSF of China (Grant No. 10605013),
Program for New Century Excellent Talents in University, and the
Project-sponsored by SRF for ROCS, SEM.

\end{document}